\documentclass[12pt]{article}

\usepackage{setspace}
\usepackage{hyperref}
\usepackage{url}
\usepackage{amssymb}
\usepackage{amsmath}
\usepackage{a4wide}
\usepackage{cancel}
\usepackage[symbol]{footmisc}

\usepackage{subcaption}
\usepackage{graphicx}
\usepackage{color}
\usepackage{lineno}

\usepackage[document]{ragged2e}

\newcommand{\labeq}[1]{\begin{equation} #1 \end{equation}}
\newcommand{\labalign}[1]{\begin{align} #1 \end{align}}

\newcommand{ \ee }{ \text{e} }
\newcommand{ \ph }{ \text{ph} }

\newcommand{ \mb }{ \mathbf } 
\newcommand{ \qq }{ \mb{q}} 
\newcommand{ \kk }{ \mb{k} } 
\newcommand{ \vv }{ \mb{v} }

\newcommand{ \fdsub }[1]{ f^{0}_{#1} }
\newcommand{ \besub }[1]{ n^{0}_{#1} }
\newcommand{ \velsub }[1]{ \mb{v}_{#1} }

\newcommand{ \gksq }{ \left | g^{smn}_{\langle\kk\rangle\qq} \right |^{2}  }
\newcommand{ \gqsq }{ \left | g^{smn}_{\kk\langle\qq\rangle} \right |^{2}  }
\newcommand{ \ibzk }{ \langle\kk\rangle }
\newcommand{ \ibzq }{ \langle\qq\rangle }
\newcommand{ \elen }[1]{ \epsilon_{#1}  }
\newcommand{ \phen }[1]{ \hbar\omega_{#1}  }
\newcommand{ \Isub }[1]{ \mb{I}_{#1} }
\newcommand{ \Fsub }[1]{ \mb{F}_{#1} }
\newcommand{ \Jsub }[1]{ \mb{J}_{#1} }
\newcommand{ \Gsub }[1]{ \mb{G}_{#1} }
\newcommand{\stirling}[2]{\genfrac{\{}{\}}{0pt}{0}{#1}{#2}}
\newcommand{\ket}[1]{| #1 \rangle}
\newcommand{\bra}[1]{\langle #1 |}

\newcommand{\kB}{k_{\text{B}}}

\begin{document}

\begin{center}
{\huge{The \texttt{elphbolt} \textit{ab initio} solver for the coupled electron-phonon Boltzmann transport equations}}
\\~\\
Nakib H. Protik$^{a}$ \footnote{Corresponding author: nakib.haider.protik@gmail.com}, Chunhua Li$^{b}$, Miguel Pruneda$^{a}$,
David Broido$^{b}$, Pablo Ordej\'on$^{a}$ \footnote{Corresponding author: pablo.ordejon@icn2.cat} \\~\\
$^{a}$\textit{\small Catalan Institute of Nanoscience and Nanotechnology (ICN2), CSIC and BIST, Campus Bellaterra, 8193 Barcelona, Spain} \\
$^{b}$\textit{\small Department of Physics, Boston College, Chestnut Hill, Massachusetts 02467, USA}
\end{center}

\newpage

\hrule
\begin{abstract}
\texttt{elphbolt} is a modern Fortran (2018 standard) code for efficiently solving the coupled electron-phonon Boltzmann transport equations from first principles. Using results from density functional and density functional perturbation theory as inputs, it can calculate the effect of the non-equilibrium phonons on the electronic transport (phonon drag) and non-equilibrium electrons on the phononic transport (electron drag) in a fully self-consistent manner and obeying the constraints mandated by thermodynamics. It can calculate the lattice, charge, and thermoelectric transport coefficients for the temperature gradient and electric fields, and the effect of the mutual electron-phonon drag on these transport properties. The code fully exploits the symmetries of the crystal and the transport-active window to allow the sampling of extremely fine electron and phonon wave vector meshes required for accurately capturing the drag phenomena. The \texttt{coarray} feature of modern Fortran, which offers native and convenient support for parallelization, is utilized. The code is compact, readable, well-documented, and extensible by design.
\end{abstract}
\hrule

\begin{center}
\end{center}

\newpage

\doublespacing

\section*{Introduction} \label{intro}
\textit{Ab initio} computation of the transport properties of materials allows a deeper understanding and probing of the rich underlying physics. It is also crucial for the predictive designing of materials for industrial applications. The advances in density functional theory (DFT) \cite{hohenberg1964inhomogeneous, kohn1965self} and density functional perturbation theory (DFPT) \cite{baroni2001phonons} have enabled accurate electron (e) and phonon (ph) band structure calculations. Furthermore, various freely available codes exist that allow fast computation of ph-ph \cite{li2014shengbte} and e-ph \cite{ponce2016epw, zhou2021perturbo} interactions. On the transport side, significant progress has been made that allows computations of the mode (band/branch and wave vector) resolved lifetimes of electrons and phonons. The phonon thermal and the electronic charge conductivity can now be calculated \textit{ab initio} in the relaxation time approximation (RTA) or, going one step further, from a full solution of the corresponding single-species Boltzmann transport equation (BTE) \cite{broido2007intrinsic, li2014shengbte, liu2017first, ponce2018towards, zhou2021perturbo}. In an interacting e-ph gas, however, the transport of the two systems is intimately connected, and a unified, coupled e-ph transport may, instead, be needed for an accurate description of the underlying physics.

The idea of a complete separation of the e and ph BTEs dates back to the 1930's and is known as Bloch's Assumption \cite{bloch1930elektrischen}. A paradoxical consequence of this assumption is that when one solves the e (ph) BTE, the phonon (electron) system is taken to remain in equilibrium. This framework was promptly questioned by Peierls \cite{peierls1930theorie}, who argued that there must exist a momentum-mixing between the electrons and the phonons, which would, in turn, cause both systems to move under the influence of a driving field. A theory of the coupled e-ph transport was developed by Gurevich in 1946 \cite{gurevich1989electron}, which allowed calculation of the effect of the non-equilibrium electrons (phonons) on the phonons (electrons). The first is called the electron drag and the latter, phonon drag. In an interacting electron-phonon gas, however, the two effects are inseparable and a mutual electron-phonon drag effect occurs. A decade later the first experimental evidence for the drag effect on the thermopower of germanium and silicon was found by Frederikse \cite{frederikse1953thermoelectric} and Geballe and Hull \cite{geballe1954seebeck, geballe1955seebeck}. Around the same time, an influential theory was developed by Conyers Herring \cite{herring1954theory} to explain the experimental findings. Herring's theory, which contains several free-parameters, partially decouples the e and the ph BTEs, retaining an approximate term to account for the drag effect. From a physical point of view, however, Herring's theory violates a deep, thermodynamic connection between the Seebeck and the Peltier effects knows as the Kelvin-Onsager relationship \cite{sondheimer1956kelvin}. Semi-analytical work on 2D systems was carried out in 1987 by Cantrell, Butcher, and coworkers \cite{cantrell1987I, cantrell1987II}. More recently, in 2014, a semi-analytical model with \textit{ab initio} fitted parameters and a partially decoupled framework was employed by Mahan, Lindsay, and Broido to calculate the drag effect on the thermopower of silicon \cite{mahan2014seebeck}. Soon after, fully \textit{ab initio} drag calculations with partially decoupled solutions of the e and ph BTEs were devised by Zhou et. al. in 2015 for silicon \cite{zhou2015ab}. The code used from that work was released in 2020 \cite{zhoucode2020}. \textit{Ab initio} solutions to partially decoupled e and ph BTEs were also developed by Fiorentini and Bonini in 2016 for silicon \cite{fiorentini2016thermoelectric} and Macheda and Bonini \cite{macheda2018magnetotransport} in 2018 for diamond. Finally, in 2020, a fully coupled e-ph BTEs solution was devised and applied to gallium arsenide using model e-ph interactions by Protik and Broido \cite{protik2020coupled}. In the same year, that method was improved to include fully \textit{ab initio} e-ph interactions to calculate the drag effect in silicon carbide by Protik and Kozinsky \cite{protik2020electron}. 

Here we present \texttt{elphbolt} (short for \textbf{el}ectron-\textbf{ph}onon \textbf{Bol}tzmann \textbf{t}ransport), a code that features major improvements over the methods given in Refs. \cite{protik2020coupled, protik2020electron}. Moreover, we release the code as Free/Libre software under a GNU General Public License version 3, bringing the capabilities for calculating the e-ph drag physics via an \textit{ab initio} solution of the fully coupled e-ph BTEs, almost a century after Peierls' conception of the idea, to the broader transport physics community. Our code is hosted on github \cite{protik_elphbolt_2021}.
\begin{figure}
	\centering
	\includegraphics[width=0.7\linewidth]{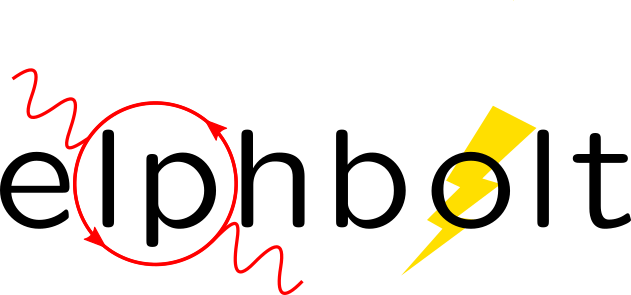}
	\caption{The logo of \texttt{elphbolt}. The red symbol is a phonon self-energy diagram and the yellow bolt signifies both the electron transport and the efficiency of the code.}
	\label{fig:elphboltlogo}
\end{figure}

Using an \textit{ab initio} and Kelvin-Onsager relationship conserving solution of the coupled e-ph BTEs, \texttt{elphbolt} gives access to the:
\begin{itemize}
	\item mode resolved phonon thermal conductivity;
	\item mode resolved phonon Peltier coefficient;
	\item mode resolved electronic charge conductivity;
	\item mode resolved electronic thermal conductivity;
	\item mode resolved electronic Seebeck and Peltier coefficients;
	\item and the effect of e-ph drag on all of the above quantities.
\end{itemize}

This code is suitable for the study of 3d and 2d insulators, semiconductors, semimetals, and metals.

In the sections below, we present the theory behind \texttt{elphbolt}, the implementation, and outlook.

\section*{Results} \label{results}
In this section, we give results for the calculated thermopower, mobility, and thermal conductivity of \textit{n}-doped silicon. First, the basic ingredients of the theory are described in detail. We present the elementary interactions considered in this work. This is followed by a discussion of the BTEs in the forms in which they are implemented, and the various types of solutions that \texttt{elphbolt} offers. Lastly, we present the transport coefficients that are obtained from the solution of the BTEs along with a brief discussion of the Kelvin-Onsager reciprocal relationship connecting the thermoelectric coefficients. 
\subsection*{Electrons, phonons, and interactions} \label{interact}
First, we need the electrons and phonons, and their interactions calculated on arbitrarily fine wave vector meshes. These are achieved by the Wannier interpolation techniques described in Refs. \cite{marzari1997maximally, souza2001maximally, giustino2007electron, ponce2016epw}. We refer the readers to these seminal works for the details of the calculation of the Wannier functions and obtaining the Wannier representations of the various physical quantities starting from their Bloch representations. In particular, the expressions for the Hamiltonian, dynamical matrix, and the e-ph matrix elements are given in Refs. \cite{giustino2007electron, ponce2016epw}. Below we simply provide the expressions for the Wannier to Bloch transformations that are computed within \texttt{elphbolt}.

The Hamiltonian in the Bloch representation can be obtained by the Fourier transformation:
\labeq{
	H_{mn}(\kk) = \dfrac{1}{N_{\ee}}\sum_{\mb{R}_{\ee}}\exp(i\kk\cdot\mb{R}_{\ee})H_{mn}(\mb{R}_{\ee}),
}
where $m$ and $n$ are band indices, $\kk$ is an arbitrary wave vector, $H(\mb{R}_{\ee})$ is the Hamiltonian in the Wannier representation, living on the real space spanned by $\{\mb{R}_{\ee}\}$, and $N_{\ee}$ is the number of real space unit cells.

Diagonalizing $H(\kk)$, we obtain the electronic band energies $\elen{m\kk}$. The unitary matrix $U_{\kk}$ digonalizing the Hamiltonian contains the eigenstates $\ket{m\kk}$. The band velocities are obtained from the Hellmann-Feynman theorem \cite{feynman1939forces}:
\labeq{
	\vv_{m\kk} = \dfrac{1}{\hbar}\bra{m\kk}\nabla_{\kk}H(\kk)\ket{m\kk},
}
where $\hbar$ is the reduced Planck constant.

Similarly, the dynamical matrix in the Bloch representation is obtained from its Wannier representation using
\labeq{
	D_{ss'}(\qq) = \dfrac{1}{N_{\ph}}\sum_{\mb{R}_{\ph}}\exp(i\qq\cdot\mb{R}_{\ph})D_{ss'}(\mb{R}_{\ph}) + D^{\text{NAC}}_{ss'}(\qq),
}
where $s$ and $s'$ denote the phonon branches, $\qq$ is an arbitrary wave vector, $D_{ss'}(\mb{R}_{\ph})$ is the dynamical matrix in the Wannier representation, and $\mb{R}_{\ph}$ locates each unit cell in real space containing $N_{\ph}$ cells. The first term is short-ranged, and the term $D^{\text{NAC}}_{ss'}$ is the long-range (the, so called, non-analytic) correction due to the dipole-dipole interaction given by \cite{pick1970microscopic}
\labeq{
	D^{\text{NAC}}_{ss'}(\qq) = \dfrac{1}{\sqrt{m_{\tau}m_{\tau'}}}\dfrac{e^{2}}{\varepsilon_{0}V}\dfrac{\qq\cdot Z^{*}_{\tau}\otimes\qq\cdot Z^{*}_{\tau'}}{\qq\cdot\epsilon^{\infty}\cdot\qq},
}
where $e$ is the electronic charge, $V$ is the primitive unit cell volume, $\tau$ and $\tau'$ label the basis atoms with masses $m_{\tau}$ and $m_{\tau'}$, $Z^{*}$ is the Born effective charge tensor, $\varepsilon_{0}$ is the permittivity of free space, and $\epsilon^{\infty}$ is the high-frequency dielectric tensor. This additional term is only required for polar materials.

We obtain the phonon branch energies $\hbar\omega_{s\qq}$ and the eigenstates $\ket{s\qq}$ by diagonalizing $D_{ss'}(\qq)$ with the unitary matrix $u_{\qq}$. Here $\omega_{s\qq}$ is the phonon angular frequency. The phonon group velocities are calculated using
\labeq{
	\vv_{s\qq} = \dfrac{1}{2\omega_{s\qq}}\bra{s\qq}\nabla_{\qq}D(\qq)\ket{s\qq}.
}

Equipped with these electronic and phononic quantities, we move on to the calculation of the various interactions in the electron-phonon system.

We start with the e-ph interactions. The vertices (matrix elements) in Bloch space are given by \cite{giustino2007electron}
\labeq{
	g^{smn}_{\kk\qq} = \sqrt{\dfrac{\hbar}{2m_{\tau}\hbar\omega_{s\qq}}}\dfrac{1}{N_{\ee}N_{\text{ph}}}\sum_{\mb{R}_{\ee}\mb{R}_{\text{ph}}}\exp(i\kk\cdot\mb{R}_{\ee} + i\qq\cdot\mb{R}_{\text{ph}})\sum_{s'm'n'}g^{s'm'n'}_{\mb{R}_{\ee}\mb{R}_{\text{ph}}}U_{nn'\kk'}U^{\dagger}_{m'm\kk}u_{s's\qq} + g^{\text{Fr}, smn}_{\kk\qq},
}

where $g^{smn}_{\mb{R}_{\ee}\mb{R}_{\text{ph}}}$ are the e-ph matrix elements in Wannier space. The additional final term is the long-range correction due to the so-called Fr\"{o}hlich interaction \cite{sjakste2015wannier,verdi2015frohlich} required for polar materials. For these materials, the $g_{\mb{R}_{\ee}\mb{R}_{\text{ph}}}$ is constructed to be the short-range part of the full interaction. The Fr\"{o}hlich term takes the following form:

\labalign{
	g^{\text{Fr}, smn}_{\kk\qq} = i\dfrac{ e^{2}}{\varepsilon_{0}V}\sum_{\tau}\sqrt{\dfrac{\hbar}{2m_{\tau}\hbar\omega_{s\qq}}} \Big[U_{\kk'}U^{\dagger}_{\kk}\Big]_{nm} \sum_{\mathcal{G} \neq -\qq} &\dfrac{(\qq+\mathcal{G})\cdot Z^{*}_{\tau}\cdot \mb{\xi}_{\tau,s\qq}}{(\qq+\mathcal{G})\cdot\epsilon^{\infty}\cdot(\qq+\mathcal{G})} \\ \nonumber
	&\times \exp\left[-i(\qq + \mathcal{G})\cdot \mb{r}_{\tau}\right],
}
where $\mb{\xi}_{\tau,s\qq} = u_{\tau,s\qq}$ is the eigendisplacement of the basis atom $\tau$ due to the phonon mode $s\qq$, $\mb{r}_{\tau}$ is the position of the atom $\tau$ in the primitive unit cell, and the sum over $\mathcal{G}$ is performed using the Ewald sum technique by multiplying each term of the summation by the factor $\exp[-(\qq+\mathcal{G})\cdot\epsilon^{\infty}\cdot(\qq+\mathcal{G})/(4\alpha)]$, where the parameter $\alpha$ is set to $1$ Bohr$^{-2}$.

Recently, it has been shown that quadrupolar corrections may be important for an accurate description of the e-acoustic phonon interactions \cite{brunin2020electron, jhalani2020piezoelectric}. Particularly, the electronic mobility can be strongly affected for those piezoelectric materials in which the transport active band extremum is at the BZ center. Currently, we do not have the capabilities for including the quadrupolar corrections. As such, care must be taken when dealing with strongly polar materials such as cubic GaAs and wurtzite GaN that feature zone-centered band extrema. We plan to include support for quadrupolar corrections in a future release of the code.

In terms of the e-ph matrix elements described above, the temperature dependent transition rates of the electrons due to the phonon absorption (+) and emission (-) processes are given by \cite{smithtransport1989}

\begin{align}\label{eq:Xeph}
	\stirling{X^{\text{e-ph},+}_{m\ibzk n\kk'|s\qq}}{X^{\text{e-ph},-}_{m\ibzk n\kk'|s\qq}} = \dfrac{2\pi}{\hbar N_{\kk}}\gksq\stirling{ \fdsub{m\ibzk}(1-\fdsub{n\kk'})\besub{s\qq}\delta(\elen{n\kk'}-\elen{m\ibzk}-\phen{s\qq}) } { \fdsub{m\ibzk}(1-\fdsub{n\kk'})(1+\besub{s-\qq})\delta(\elen{n\kk'}-\elen{m\ibzk}+\phen{s-\qq}) },
\end{align}

where $f^{0}$ and $n^{0}$ is the equilibrium, i.e. the Fermi (Bose), distribution of the electron (phonon) gas, and the delta functions enforce the energy conservation in a scattering process. The notation $\ibzk\kk'|\qq$ above means that the initial electron wave vector is taken to be in the irreducible Brillouin zone (IBZ), the final electron wave vector is on the full first Brillouin zone (FBZ), the phonon wave vector mediating this transition is $\qq = [\kk' - \ibzk]$, with the square brackets denoting a modulo operation with a reciprocal lattice vector. $N_{\kk}$ is the number of electronic wave vectors in the FBZ.
 
In \texttt{elphbolt}, we have the option of including electron-charged impurity (e-chimp) scattering to capture the effect of charged dopants on the transport properties. The e-chimp interaction is calculated within the first Born approximation, taking the impurity potential to have a static, Yukawa (screened Coulomb) form. The modulus squared of the interaction vertex is given by (generalized from Ref. \cite{chattopadhyay1981electron})
\labeq{
	\left|g^{\text{e-chimp}}_{\qq}\right|^{2} = \dfrac{1}{V}\sum_{i}n_{i}\left[\dfrac{Z_{i}e}{\varepsilon_{0}\epsilon^{0}(q^{2} + q_{\text{TF}}^{2})}\right]^{2}, 
}
where $i = {p, n}$ denotes $p$- or $n$-type doping, $n_{i}$ is the concentration of the dopant, $Z_{i}$ is the ionization of the dopant, $e$ is the absolute electronic charge, $\epsilon^{0}$ is the zero frequency dielectric constant of the material, $q$ is the wave vector magnitude measured from the nearest BZ center, and $q_{\text{TF}}$ is the Thomas-Fermi screening wave vector given by
\labeq{
	q_{\text{TF}}^{2} = \dfrac{d_{s}e^{2}\beta}{N_{\kk}V\varepsilon_{0}\epsilon^{0}}\sum_{m\kk}f^{0}_{m\kk}\left(1 - f^{0}_{m\kk}\right),
}
where $d_{s}$ is the spin degeneracy of the electronic state, and $\beta \equiv (\kB T)^{-1}$ in terms of the Boltzmann constant $\kB$ and temperature $T$.

In terms of the matrix elements, the charged impurity mediated $m\ibzk$ to $n\kk'$ transition rates are given by
\labeq{
	X^{\text{e-chimp}}_{m\ibzk n\kk'|\qq} = \dfrac{2\pi}{\hbar N_{\kk}} \left|g^{\text{e-chimp}}_{\qq}\right|^{2}\left(1 - \dfrac{\ibzk\cdot \kk'}{kk'}\right) f^{0}_{m\ibzk}\left(1-f^{0}_{m\ibzk}\right)\delta(\epsilon_{n\kk'} - \epsilon_{m\ibzk}),
}
where we have multiplied the squared matrix elements with a transport factor corresponding to the in-scattering for elastic processes. Note that this term suppresses forward scattering.

At the present, we do not include e-e interactions. Typically, the e-e scattering does not strongly affect the transport in metals \cite{smithtransport1989}. However, in degenerate semiconductors, its effects might be strong \cite{caruso2016theory}. A rigorous treatment of the e-e interaction necessarily requires the calculation of the dynamical screening of the electron gas. Typically, this is done in the random phase approximation which introduces an additional Bosonic system - the plasmons. We wish to include e-e and the related e-plasmon scattering in a future release.

Next, we look at the interactions of the phonons. We start with the ph-e interaction. The lowest-order process involves two electrons, and the transition probabilities are given by \cite{smithtransport1989}
\labeq{
	Y^{\text{ph-e}}_{s\ibzq|m\kk n\kk'} = \dfrac{2\pi}{\hbar N_{\qq}}\gqsq \fdsub{m\kk}(1-\fdsub{n\kk'})\besub{s\ibzq}\delta(\elen{n\kk'}-\elen{m\kk}-\phen{s\ibzq}),
}
where $N_{\qq}$ is the number of phonon wave vectors in the FBZ and $\kk' = [\kk + \ibzq]$. This describes the same + process given in Eq. \eqref{eq:Xeph}.

The lowest-order ph-ph interaction vertices describing the $\qq'' \rightarrow \qq \pm \qq'$ processes are given by \cite{li2014shengbte}
\labeq{
	V^{\pm,ss's''}_{\ibzq\qq'\qq''} = \sum_{i}^{'}\sum_{jk}\sum_{\alpha\beta\gamma}\Psi^{\alpha\beta\gamma}_{ijk}\dfrac{e^{\alpha}_{i,s\ibzq}e^{\beta}_{j,s'\pm\mathbf{q}'}e^{\gamma}_{k,s''-\mathbf{q}''}}{\sqrt{m_{i}m_{j}m_{k}}},
}
where $i, j, k$ label the atoms in the supercell, the primed sum indicates restriction to the central primitive unit cell, the Greek letters are Cartesian directions, $\mb{e}_{j,s'\qq'} = \exp(i\mb{r}_{j}\cdot{\qq'})\mb{\xi}_{j,s'\qq'}$ is the eigendisplacement of atom $j$ in the supercell due to the phonon mode $s'\qq'$, and $\Psi_{ijk}$ are the third order anharmonic force constants.

We will require only the modulus squared of the above expression. As such, we need only calculate the $V^{-}$ processes, since $\left|V^{+,ss's''}_{\qq\qq'\qq''}\right|^{2} = \left|V^{-,ss's''}_{\qq-\qq'\qq''}\right|^{2}$.

In terms of the ph-ph vertices, the anharmonic phonon transition rates can be calculated as \cite{li2014shengbte}

\labalign{
	W^{\pm}_{s\ibzq s'\qq'|s''\qq''} = &\dfrac{\pi\hbar}{4N_{\qq}}\dfrac{\left|V^{\pm,ss's''}_{\ibzq\qq'\qq''}\right|^{2}}{\omega_{s\ibzq}\omega_{s'\qq'}\omega_{s''\qq''}} \\ \nonumber 
	&\times (\besub{s\ibzq}+1)\left(\besub{s\qq'}+\dfrac{1}{2} \pm \dfrac{1}{2}\right)\besub{s\qq''} \delta(\omega_{s\ibzq}\pm\omega_{s\qq'}-\omega_{s\qq''}).
}

We note that four-phonon interactions have been shown to be important for phonon transport in strongly anharmonic materials and those that feature anomalously weak three-phonon scattering rates for modes that dominate the transport at high temperatures; see Ref. \cite{feng2020higher} and the references therein. We do not currently have the functionality for calculating four-phonon interactions but plan to include it in a future release.

In \texttt{elphbolt}, we may also consider lowest-order ph-isotope and ph-substitution defect interactions within the first Born approximation and assuming the defect to be an on-site perturbation and low in concentration. This approximation can be described as the Tamura model \cite{tamura1983isotope}. The transition rates for these two phonon processes are given by
\labeq{
	W^{\text{ph-x}}_{s\ibzq s'\qq'} = \dfrac{\pi\omega^{2}_{s\ibzq}}{2}n^{0}_{s\ibzq}(1 + n^{0}_{s\qq})\sum_{\tau}g_{\tau}\left|\xi^{*}_{\tau,s\ibzq}\cdot \xi_{\tau,s'\qq'}\right|^{2}\delta(\omega_{s'\qq'} - \omega_{s\ibzq}), 
}
where $x = $ \{\text{iso, subs}\} for isotope and substitution defect scattering, respectively, and
\labeq{
	g_{\tau} = \sum_{t}f_{t\tau}\left(1 - \dfrac{M_{t\tau}}{\langle M\rangle_{\tau} } \right)^{2}
}
is the mass variance parameter of the host atom $\tau$, $t$ denotes the type -- isotope or substitution -- of the guest atom, $f_{t\tau}$ is the ratio of type-$t$ guest to type-$\tau$ host, and $\langle M\rangle_{\tau}$ is the average on-site mass.

It is worth noting that the above description of the ph-defect interaction is simplistic and can fail for defects which cause extended bond perturbations. There exist advanced diagrammatic techniques that can handle such cases better \cite{mingo2010cluster, katcho2014effect, fava2021dopants}. Such methods are planned for inclusion in a future release of \texttt{elphbolt}.

\subsection*{Coupled and decoupled BTEs} \label{btes}
In \texttt{elphbolt} we may consider the electric ($\mb{E}$) and temperature gradient ($\mb{\nabla} T$) fields as the drivers of the electron and the phonon currents. These applied fields cause the distribution functions of the electrons and the phonons, $f_{m\kk}$ and $n_{s\qq}$, respectively, to deviate from their equilibrium forms. In the linear response regime, these are given by
	
\labalign{\label{eq:devfns}
	f_{m\kk} &\approx f^{0}_{m\kk}\left[ 1 + (1 - f^{0}_{m\kk})\Psi_{m\kk} \right] \nonumber \\
	n_{s\qq} &\approx n^{0}_{s\qq}\left[ 1 + (1 + n^{0}_{s\qq})\Phi_{s\qq} \right],
}

The deviation functions above of the electrons and the phonons, respectively, can be written as

\labalign{\label{eq:resfns}
	\Psi_{m\kk} &= -\beta\mb{\nabla}T\cdot \mb{I}_{m\kk} - \beta\mb{E}\cdot \mb{J}_{m\kk} \nonumber \\
	\Phi_{s\qq} &= -\beta\mb{\nabla}T\cdot \mb{F}_{s\qq} - \beta\mb{E}\cdot \mb{G}_{s\qq},
}

where $\mb{I}_{m\kk} (\mb{F}_{s\qq})$ is the electron (phonon) response function that measures the deviation from equilibrium of the occupation of the electron (phonon) state $m\kk$ ($s\qq$) due to applied $\mb{\nabla}T$ field. Similarly, $\mb{J}_{m\kk} (\mb{G}_{s\qq})$ is the electron (phonon) response to the $\mb{E}$ field.

The coupled electron and phonon BTEs have been given in numerous references e.g. \cite{zhou2015ab, fiorentini2016thermoelectric, macheda2018magnetotransport, protik2020coupled, protik2020electron, smithtransport1989}. Here, we write them in the following form in terms of the response functions:

\labalign{\label{eq:cbtes}
	\mb{\nabla}T \text{ field: }& \nonumber \\
	\text{e: }& \mb{I}_{m\kk} = \mb{I}^{0}_{m\kk} + \Delta\mb{I}^{\text{S}}_{m\kk}[\mb{I}_{m\kk}] + \Delta\mb{I}^{\text{D}}_{m\kk}[\mb{F}_{s\qq}] \nonumber \\
	\text{ph: }& \mb{F}_{s\qq} = \mb{F}^{0}_{s\qq} + \Delta\mb{F}^{\text{S}}_{s\qq}[\mb{F}_{s\qq}] + \Delta\mb{F}^{\text{D}}_{s\qq}[\mb{I}_{m\kk}] \nonumber \\
	\mb{E} \text{ field: }& \nonumber \\	
	\text{e: }& \mb{J}_{m\kk} = \mb{J}^{0}_{m\kk} + \Delta\mb{J}^{\text{S}}_{m\kk}[\mb{J}_{m\kk}] + \Delta\mb{J}^{\text{D}}_{m\kk}[\mb{G}_{s\qq}] \nonumber \\
	\text{ph: }& \mb{G}_{s\qq} = \cancelto{0}{\mb{G}^{0}_{s\qq}} +   \Delta\mb{G}^{\text{S}}_{s\qq}[\mb{G}_{s\qq}] + \Delta\mb{G}^{\text{D}}_{s\qq}[\mb{J}_{m\kk}]. \nonumber \\
}

In the equations above, the terms with the superscript $0$ are the relaxation time approximation (RTA) terms. These involve a direct coupling to the applied field, which is why in the phonon equation to the $\mb{E}$ field, the RTA term is identically zero. Furthermore, the RTA only involves out-scattering processes. Truncating the BTEs up to the RTA term is equivalent to ignoring the in-scattering corrections and the drag effect. Next, the in-scattering corrections are given by the terms with the superscript S. These terms are functionals of the response function of the same species and, as such, they are called the self terms. The inclusion of these terms renders the BTEs (barring the one for $\mb{G}$) into a set of decoupled equations which can be solved iteratively. Truncating the BTEs up to the self term, however, still ignores the drag effect. This is because, at this level of the approximation, the electron equations implicitly take phonons to remain in equilibrium, while the phonon equations take electrons to remain in equilibrium. Finally, the terms with the superscript D represent the drag terms. These are functionals of the other species. Inclusion of these terms allows an accurate description of the mutual drag effect in the interacting electron-phonon system. At this level, each pair of BTEs must be solved self-consistently. Furthermore, there exists a thermodynamic restriction known as the Kelvin-Onsager relationship \cite{sondheimer1956kelvin} that reflects a deeper coupling of the two pairs of equations for the two fields. In \texttt{elphbolt}, we have devised a fast, iterative scheme that allows us to obtain the full solution of the Eqs. \eqref{eq:cbtes}, while respecting the thermodynamic restrictions mandated by the Kelvin-Onsager relationship. This is discussed further in the \textit{Transport coefficients} subsection.

Now we provide the expressions for each term in the BTEs in Eq. \eqref{eq:cbtes}. First we give the expressions for the electronic equations. The field coupling, RTA terms are given by
\labeq{
	\stirling{\Isub{m\kk}^{0}}{\Jsub{m\kk}^{0}} = \stirling{(\elen{m\kk}-\mu_{c})/T}{e}\dfrac{\velsub{m\kk}}{W^{\text{RTA}}_{m\kk}},
}
where $\mu_{c}$ is the chemical potential of the electronic system. The electron RTA scattering rates are given by
\labeq{
	W^{\text{RTA}}_{m\kk} = \dfrac{1}{f^{0}_{m\kk}(1 - f^{0}_{m\kk})}\left[\sum_{ns\qq} \left(X^{+}_{m\kk n\kk'|s\qq} + X^{-}_{m\kk n\kk'|s\qq}\right) + \sum_{n\qq}X^{\text{e-chimp}}_{m\kk n\kk'|\qq}\right].
}
The summation above of the independent scattering channels at the RTA level is known as the Matthiessen's Rule.

The self terms are given by
\labeq{
	\stirling{\Delta \Isub{m\kk}^{\text{S}}}{\Delta \Jsub{m\kk}^{\text{S}}} = \dfrac{1}{f^{0}_{m\kk}(1 - f^{0}_{m\kk})}\dfrac{1}{W^{\text{RTA}}_{m\kk}}\sum_{sn\kk'} \stirling{\Isub{n\kk'}}{\Jsub{n\kk'}}\left(X^{+}_{m\kk n\kk'|s\qq} + X^{-}_{m\kk n\kk'|s\qq}\right).
}

Lastly, the drag terms are
\labeq{
	\stirling{\Delta \Isub{m\kk}^{\text{D}}}{\Delta \Jsub{m\kk}^{\text{D}}} = \dfrac{1}{f^{0}_{m\kk}(1 - f^{0}_{m\kk})}\dfrac{1}{W^{\text{RTA}}_{m\kk}}\sum_{sn\kk'} \stirling{ -\mb{F}_{s\qq}X^{+}_{m\kk n\kk'|s\qq} + \mb{F}_{s-\qq}X^{-}_{m\kk n\kk'|s\qq}}{-\mb{G}_{s\qq}X^{+}_{m\kk n\kk'|s\qq} + \mb{G}_{s-\qq}X^{-}_{m\kk n\kk'|s\qq}}.
}

Similarly, for the phonons we have
\labeq{
	\stirling{\Fsub{s\qq}^{0}}{\Gsub{s\qq}^{0}} = \stirling{\hbar\omega_{s\qq}/T}{0}\dfrac{\velsub{s\qq}}{W^{\text{RTA}}_{s\qq}},
}
where the phonon RTA scattering rates are given by
\labeq{
	W^{\text{RTA}}_{s\qq} = \dfrac{1}{n^{0}_{s\qq}(1 + n^{0}_{s\qq})}\left[\sum_{s'\qq' s''\qq''} \left(W^{+}_{s\qq s'\qq'|s''\qq'} + \dfrac{1}{2}W^{-}_{s\qq s'\qq'|s''\qq'}\right) +
	d_{s}\sum_{mn\kk}Y_{s\qq|m\kk n\kk'}  + \sum_{s'\qq'}W^{\text{ph-x}}_{s\qq s'\qq'}\right].
}

The self and drag terms are, respectively

\labalign{
	\stirling{\Delta \Fsub{s\qq}^{\text{S}}}{\Delta \Gsub{s\qq}^{\text{S}}} = &\dfrac{1}{n^{0}_{s\qq}(1 + n^{0}_{s\qq})}\dfrac{1}{W^{\text{RTA}}_{s\qq}} \\ \nonumber
	&\times \sum_{s'\qq's''\qq''} \left[ W^{+}_{s\qq s'\qq'|s''\qq''} \stirling{\Fsub{s''\qq''} - \Fsub{s'\qq'}}{\Gsub{s''\qq''} - \Gsub{s'\qq'}} + \dfrac{1}{2} W^{-}_{s\qq s'\qq'|s''\qq''} \stirling{\Fsub{s''\qq''} + \Fsub{s'\qq'}}{\Gsub{s''\qq''} + \Gsub{s'\qq'}}\right]
}
and
\labeq{
	\stirling{\Delta \Fsub{s\qq}^{\text{D}}}{\Delta \Gsub{s\qq}^{\text{D}}} = \dfrac{d_{s}}{n^{0}_{s\qq}(1 + n^{0}_{s\qq})}\dfrac{1}{W^{\text{RTA}}_{s\qq}}\sum_{mn\kk}Y_{s\qq|m\kk n\kk'}\stirling{\mb{I}_{n\kk'} - \mb{I}_{m\kk}}{\mb{J}_{n\kk'} - \mb{J}_{m\kk}}.
}

With these expressions, we solve the coupled e-ph BTEs, Eqs. \eqref{eq:cbtes}, using an iterative procedure which is an improvement over the one used earlier in Refs. \cite{protik2020coupled, protik2020electron}. Here we use a unified scheme, indexed by a single iterator, that of the ph BTE. In this approach, for a coupled BTEs solution, the e BTE is internally iterated to self-consistency for each iteration of the ph BTE. Thus, once the ph BTE has achieved self-consistency, the e BTE is guaranteed to do so also. This procedure is physically justified since the electron system is, in general, faster than the phonon system. Moreover, this scheme allows the strict enforcement of the Kelvin-Onsager relationship after each ph BTE iteration because the electron system is always brought to consistency with the current non-equilibrium phonon system. However, since the Kelvin-Onsager relationship mandates a thermodynamic constraint over the thermoelectric transport coefficients of both the electron and the phonon systems (see subsection \textit{Transport coefficients}), in order to confirm that this relationship is satisfied, the coupled BTEs iteration must be performed for both the $\mb{E}$ and $\mb{\nabla}T$ fields, simultaneously. Thus, all four BTEs in Eqs. \eqref{eq:cbtes} are tightly coupled. This approach is an improvement over the simple, dual iterator scheme used in Refs. \cite{protik2020coupled} and \cite{protik2020electron}, both conceptually and numerically.

We offer four different levels of solutions of the BTE for each species and field: RTA, partially decoupled, dragless full, and dragged full. Note that the RTA and dragless full solutions for the case of the phonons under the influence of the electric field are trivially zero. In our iterative approach, the computational time and space difference between the partially decoupled solution and the dragged full solution is not large, and both the RTA (iteration 0) and partially decoupled (iteration 1) solutions are provided \textit{en route} to the dragged full solution. By default, \texttt{elphbolt} will carry out a set of decoupled, hence, dragless, ph and e BTEs immediately after the coupled BTEs are fully solved. This allows the users to compare the transport coefficients in all the above mentioned approximation levels after a single computational run. The users, however, also have the option of calculating only the decoupled e and ph BTEs (dragless full), if they wish to do so.

\subsection*{Transport coefficients}
From the solutions of the response functions, we can obtain the following transport coefficient tensors.

From the electronic charge current, we get
\labeq{
\stirling{\sigma}{\sigma S} =\dfrac{d_{s}e}{V\kB T}\sum_{m\kk}f^{0}_{m\kk}(1-f^{0}_{m\kk})\mb{v}_{m\kk}\otimes  \stirling{\mb{J}_{m\kk}}{-\mb{I}_{m\kk}},
}
where $\sigma$ is the electronic charge conductivity and $S$ is the (Seebeck) thermopower.

From the electronic heat current, we get
\labeq{
	\stirling{\alpha_{\text{el}}}{\kappa_{0,\text{el}}} = -\dfrac{d_{s}}{V\kB T}\sum_{m\kk}(\epsilon_{m\kk} - \mu_{c})f^{0}_{m\kk}(1-f^{0}_{m\kk})\mb{v}_{m\kk}\otimes  \stirling{\mb{J}_{m\kk}}{-\mb{I}_{m\kk}},
}
where $\alpha_{\text{el}}$ is closely related to the electronic Peltier coefficient. The electronic component of the (Peltier) thermopower is given by $Q_{\text{el}} = \alpha_{\text{el}}(\sigma T)^{-1}$. The tensor $\kappa_{0,\text{el}}$ is the electronic thermal conductivity in the zero $\mb{E}$ field (closed circuit) condition. The electronic thermal conductivity in the open circuit condition, which is what can be measured in experiments, is given by $\kappa_{\text{el}} = \kappa_{0,\text{el}} - \alpha_{\text{el}}S$.

Lastly, from the phonon heat current, we obtain
\labeq{
	\stirling{\alpha_{\text{ph}}}{\kappa_{\text{ph}}} = \dfrac{1}{V\kB T}\sum_{s\qq}\hbar\omega_{s\qq}n^{0}_{s\qq}(1+n^{0}_{s\qq})\mb{v}_{s\qq}\otimes  \stirling{-\mb{G}_{s\qq}}{\mb{F}_{s\qq}},
}
where $\kappa_{\text{ph}}$ is the phonon thermal conductivity and $\alpha_{\text{ph}}$ is related to the phonon Peltier coefficient. Specifically, the phonon component of the (Peltier) thermopower is given by $Q_{\text{ph}} = \alpha_{\text{ph}}(\sigma T)^{-1}$.

The Kelvin-Onsager relationship unifies the Seebeck and the Peltier effects and mandates that
\labeq{
	\sigma S = \alpha T^{-1},
}
where $\alpha = \alpha_{\text{el}} + \alpha_{\text{ph}}$ \cite{sondheimer1956kelvin}. Note that this implies that the same thermopower $Q \equiv S = Q_{\text{el}} + Q_{\text{ph}}$ is obtained by both the Seebeck and the Peltier pictures.

As such, the above relationship connects the $\mb{\nabla}T$ field e BTE and the $\mb{\nabla}T$ and $\mb{E}$ field ph BTEs. Since, for a given field, the BTE for one species is coupled to the one for the other via the drag term, the Kelvin-Onsager relationship effectively leads to a tight physical connection between all the four BTEs of the interacting e-ph system. However, during the course of the iterations, numerical issues may cause the system of equations to deviate from the Kelvin-Onsager relationship. To remedy this, corrective measures must be employed. Now, within the iterative scheme described in the previous subsection, it is straightforward to enforce the Kelvin-Onsager relationship. To do this, we split $\sigma S$ into a diffusion and a drag term and a draw parallel with the Peltier decomposition:

\labalign{
	\sigma S &= \sigma S_{\text{diff}} + \sigma S_{\text{drag}} \nonumber \\
	\alpha T^{-1} &= \alpha_{\text{el}}T^{-1} + \alpha_{\text{ph}}T^{-1}.
}

Then, decomposing $\mb{I}$ as (band and wave vector indices dropped for brevity)
\labeq{
	\mb{I} = \mb{I}_{\text{diff}} + \mb{I}_{\text{drag}}
}
and demanding $\sigma S_{\text{diff}}[\mb{I}_{\text{diff}}] = \alpha_{\text{el}}T^{-1}$, we can arrive at the relation:
\labeq{
	\mb{I}_{\text{diff}} = \dfrac{\epsilon - \mu_{c}}{eT} \mb{J}.
}

The problem of enforcing the Kelvin-Onsager relationship then boils down to satisfying 
\labeq{ 
\sigma S_{\text{drag}}[\lambda \mb{I}_{\text{drag}}] = \alpha_{\text{ph}}T^{-1},
}
where $\lambda$ is a small, corrective scalar which can be easily found using a bisection method.

\subsection*{Implementation} \label{implement}
The coupled BTEs solver described above is implemented in Fortran 2018. This allows us to make use of the object-oriented programming (OOP) support and the built-in \texttt{coarray} functionality that provides concise, native syntax for parallelization. Specifically, we create the following 7 derived types: \texttt{crystal}, \texttt{symmetry}, \texttt{numerics}, \texttt{electron}, \texttt{phonon}, \texttt{epw\_wannier}, and \texttt{bte} dealing with the components of the problem that the names suggest. Each derived type contains its own data and procedures (\texttt{functions} and \texttt{subroutines}). Apart from these, there are separate \texttt{modules} for immutable \texttt{parameters}, helper procedures, etc. This hybrid OOP/procedural design enables extensibility of the code. Boilerplate \texttt{getter} and \texttt{setter} functions are generally avoided. Instead, the \texttt{intent} and \texttt{use, only} keywords of Fortran are strictly used to control the read, write, and use access of the different components of the code. This design strategy makes the code compact (about 6700 lines) for what it offers and easily readable. As a general rule, code repetition is avoided unless the generalizations lead to slow or physically unclear source. We tried to strike a balance between the speed of development, execution, readability, and extensibility. Below we discuss the general workflow and the logical structure of the program.

\subsection*{Workflow and structure}\label{workflow}
\begin{figure}
	\centering
	\includegraphics[width=1.0\linewidth]{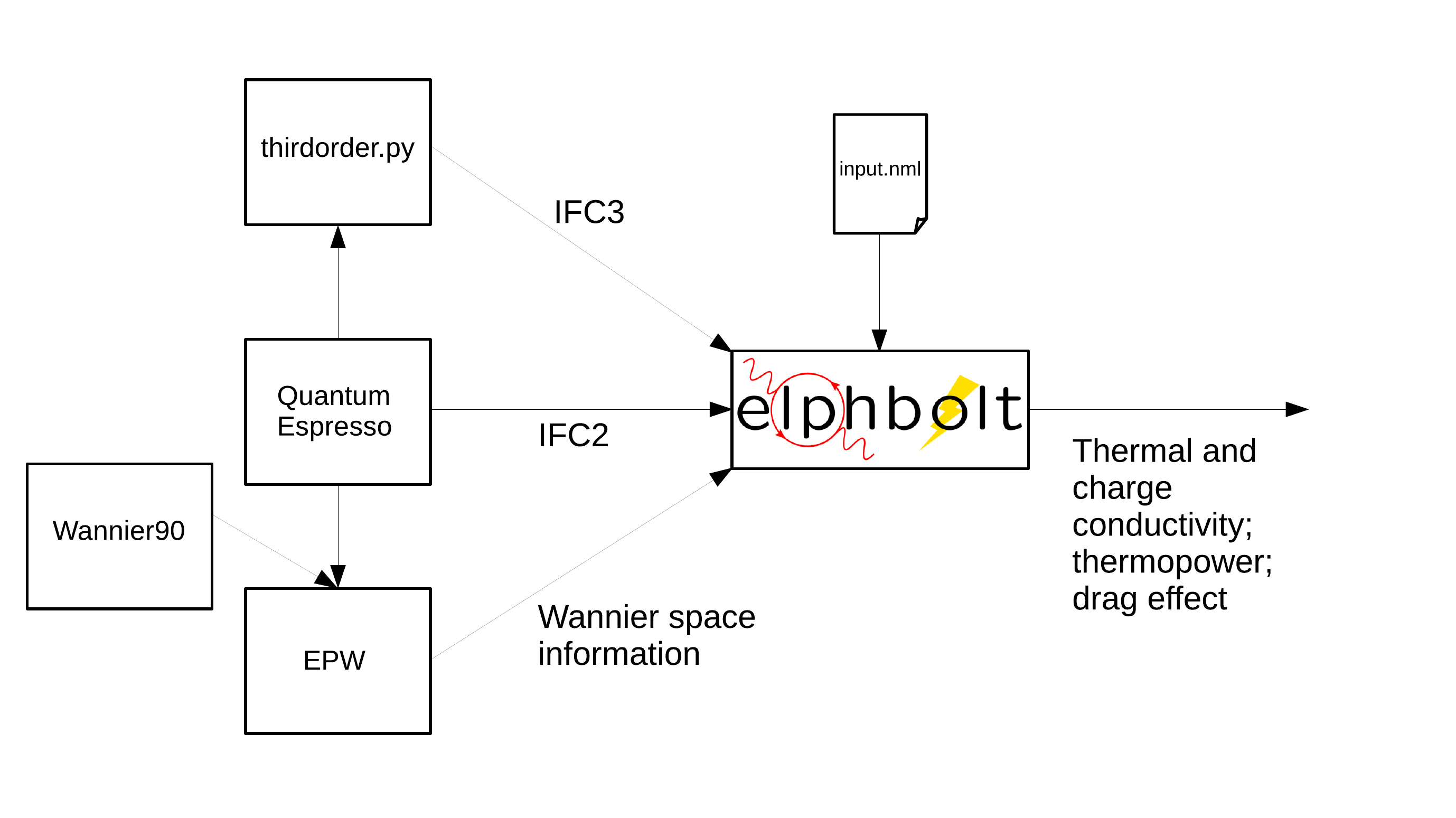}
	\caption{Workflow of \texttt{elphbolt} showing the various input data required for a coupled e-ph BTEs calculation.}
	\label{fig:workflow}
\end{figure}

In Fig. \ref{fig:workflow} we present the workflow of \texttt{elphbolt}. For a full calculation complete with the e-ph drag effect, we require the second-order interatomic force constants (IFC2s) from \texttt{Quantum Espresso} \cite{giannozzi2009quantum, giannozzi2017advanced, giannozzi2020quantum}; the third-order interatomic force constants (IFC3s) from \texttt{thirdorder.py} \cite{li2014shengbte} which provides an interface with \texttt{Quantum Espresso}; and the Wannier space electronic Hamiltonian, dynamical matrix, e-ph matrix elements, and the real space cell maps and degeneracies from \texttt{EPW} \cite{ponce2016epw, verdi2015frohlich, giustino2007electron}. \texttt{EPW} internally interfaces with \texttt{Quantum Espresso} and \texttt{Wannier90} \cite{w90}. The generation of the IFC2s and IFC3s are also part of the \texttt{ShengBTE} workflow, and the users of that code will find that only one extra step -- i.e. an \texttt{EPW} calculation -- is required to generate all the input data for an e-ph coupled BTEs calculation with \texttt{elphbolt}. We have provided two modified \texttt{EPW} source files that allow the generation of some of the required Wannier space information. The user also needs to provide an input file called \texttt{input.nml}. The format of the input is described in details in the file \texttt{README.org}.

Next, we outline the logical flow of the program. The main program is in the file \texttt{elphbolt.f90}. The program starts with creating objects of the seven derived types mentioned earlier. Then, following a welcome message, it initializes the \texttt{crystal} object. This involves reading the information about the crystal and initializing the appropriate internal variables. Following this, the \texttt{numerics} object is initialized which involves the reading in of the transport wave vector meshes, data output directory, BTE solution type, etc. Next, the \texttt{symmetry} object is initialized. At this step, the symmetries of the crystal and the BZ are calculated. Following this, the \texttt{epw\_wannier} object is initialized which involves reading in the Wannier space information. Next, we initialize the \texttt{electron} object. In this step, the information about the bands, transport active energy window, chemical potential, etc. are read in. The transport energy window restricted electronic IBZ is generated along with the bands, eigenstates, and velocities. Next we initialize the \texttt{phonon} object which involves the calculation of the phonon branches, eigenstates, and velocities. The next major step is the calculation of the interaction vertices. These zero temperature quantities are stored in the disk for later temperature and carrier concentration sweeps. The temperature dependent transition rates are calculated and stored in the disk also for their reuse in the various types of BTE solutions that are available. The transition rates expressions include the energy conserving delta functions. We provide two methods for the evaluation of these delta functions: the triangular method \cite{kurganskii1985integration, wang2017ab} and analytic tetrahedron method \cite{lambin1984computation}. The first is the only option for 2d systems and is the default for 3d systems. Unlike the more commonly used Gaussian or Lorentzian methods, the triangular and tetrahedron methods do not have any additional smearing parameter, and, as such, the electron and phonon wave vector meshes are the only parameters to converge. Finally, the \texttt{bte} object is used to solve the BTEs. The code produces a large amount of data for analysis, including the RTA scattering rates, band/branch resolved transport tensors, response functions at the RTA, partially decoupled, and fully iterated levels, among others. The users may also run the code in a post-processing mode to generate spectral transport coefficients, if required. Full descriptions of all the input options and the output data are provided in the \texttt{README.org} file.

\subsection*{Example: cubic silicon}\label{silicon}
\begin{figure}[h]
	\centering
	\includegraphics[width=0.8\linewidth]{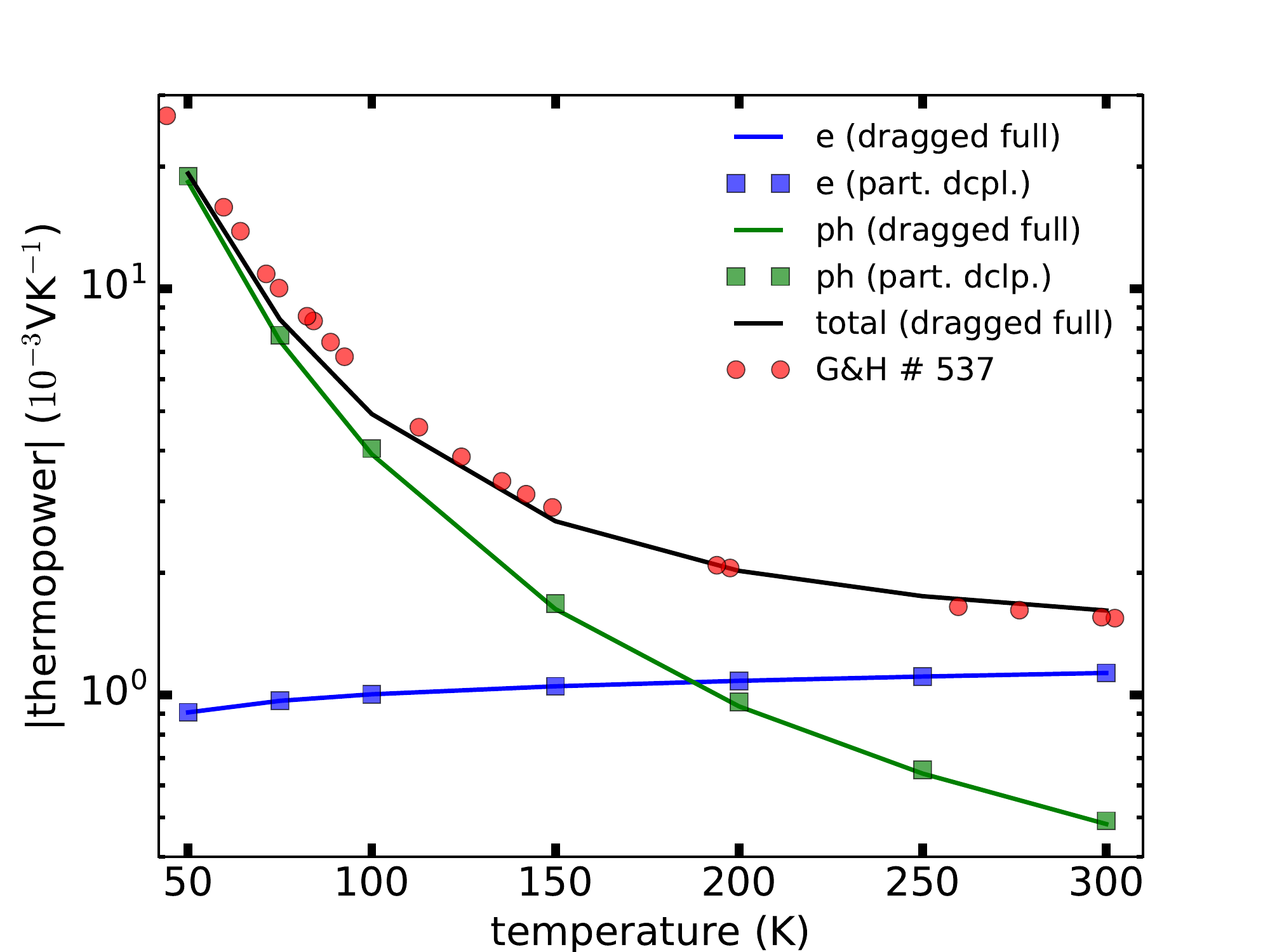}
	\caption{Temperature dependence of the thermopower of silicon for an $n$-type carrier concentration of $2.75\times 10^{14}$ cm$^{-3}$. The red circles are measurements (sample 537, concentration $2.8\times 10^{14}$ cm$^{-3}$) by Geballe and Hull \cite{geballe1955seebeck}.}
	\label{fig:lowconc_Q_T}
\end{figure}

\begin{figure}[h]
	\centering
	\includegraphics[width=0.8\linewidth]{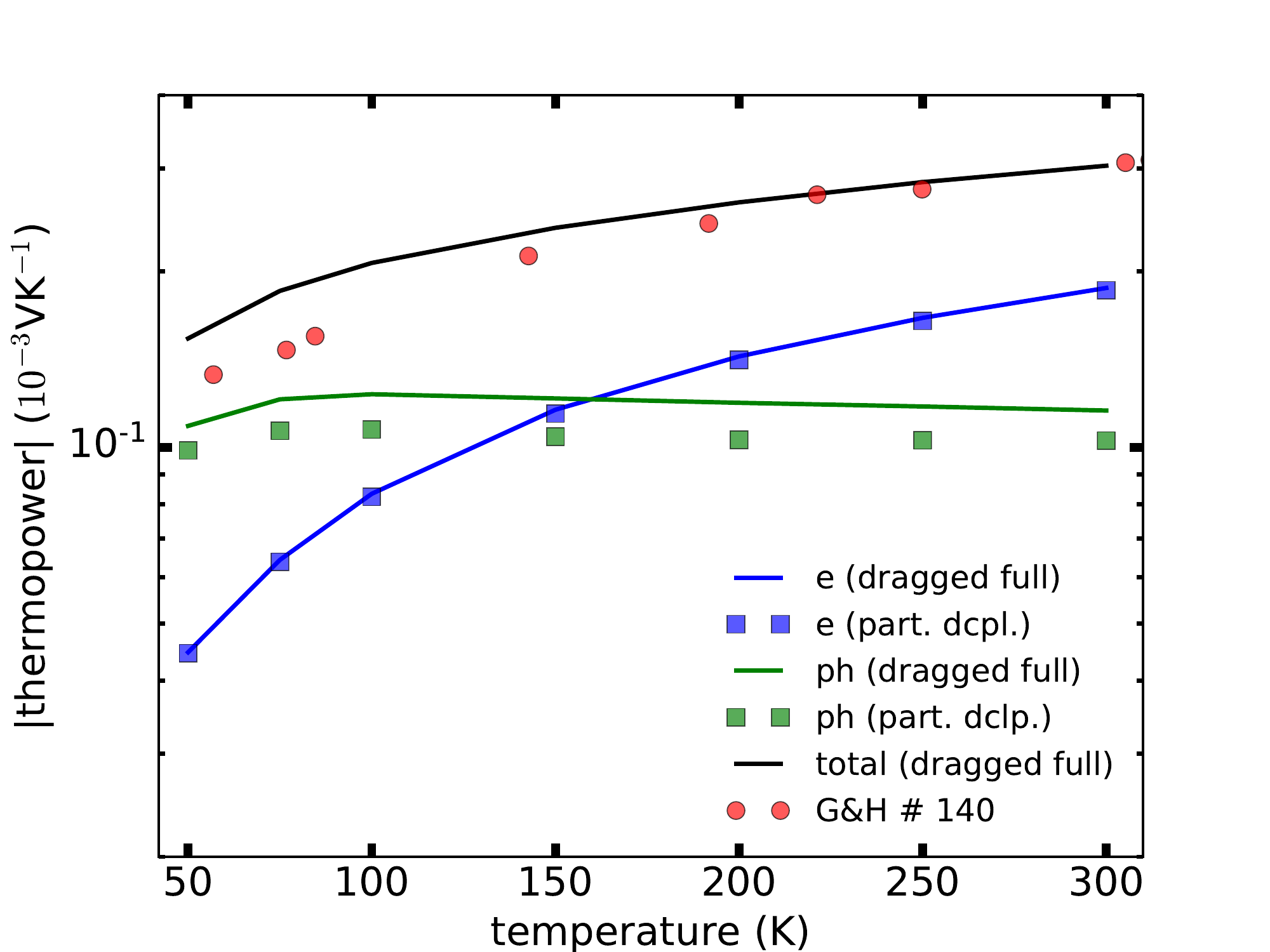}
	\caption{Temperature dependence of the thermopower of silicon for an $n$-type carrier concentration of $2.7\times 10^{19}$ cm$^{-3}$. The red circles are measurements (sample 140, concentration $2.7\times 10^{19}$ cm$^{-3}$) by Geballe and Hull \cite{geballe1955seebeck}.}
	\label{fig:highconc_Q_T}
\end{figure}

As a demonstration of the code, we calculate the drag effect on the thermopower in cubic silicon. In addition, we include calculations of the mobility and thermal conductivity. 

\begin{figure}[h]
	\centering
	\includegraphics[width=0.8\linewidth]{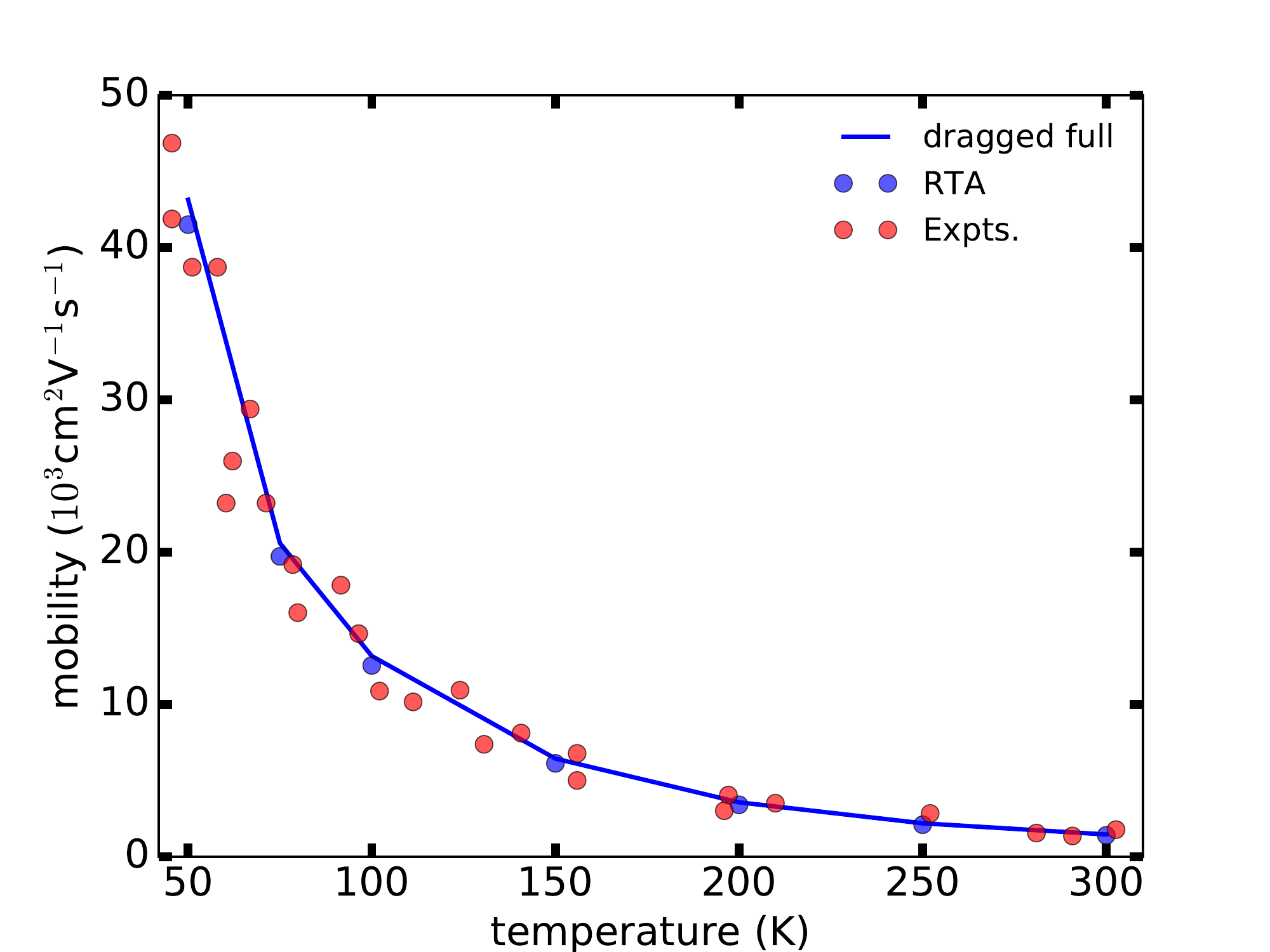}
	\caption{Temperature dependence of the mobility of silicon for an $n$-type carrier concentration of $2.75\times 10^{14}$ cm$^{-3}$. The red circles are measurements on various different samples with carrier concentrations ranging from $3.5\times 10^{13}$ to $1.4\times 10^{14}$ cm$^{-3}$ \cite{canali1975electron}.}
	\label{fig:lowconc_mob_T}
\end{figure}

\begin{figure}[h]
	\centering
	\includegraphics[width=0.8\linewidth]{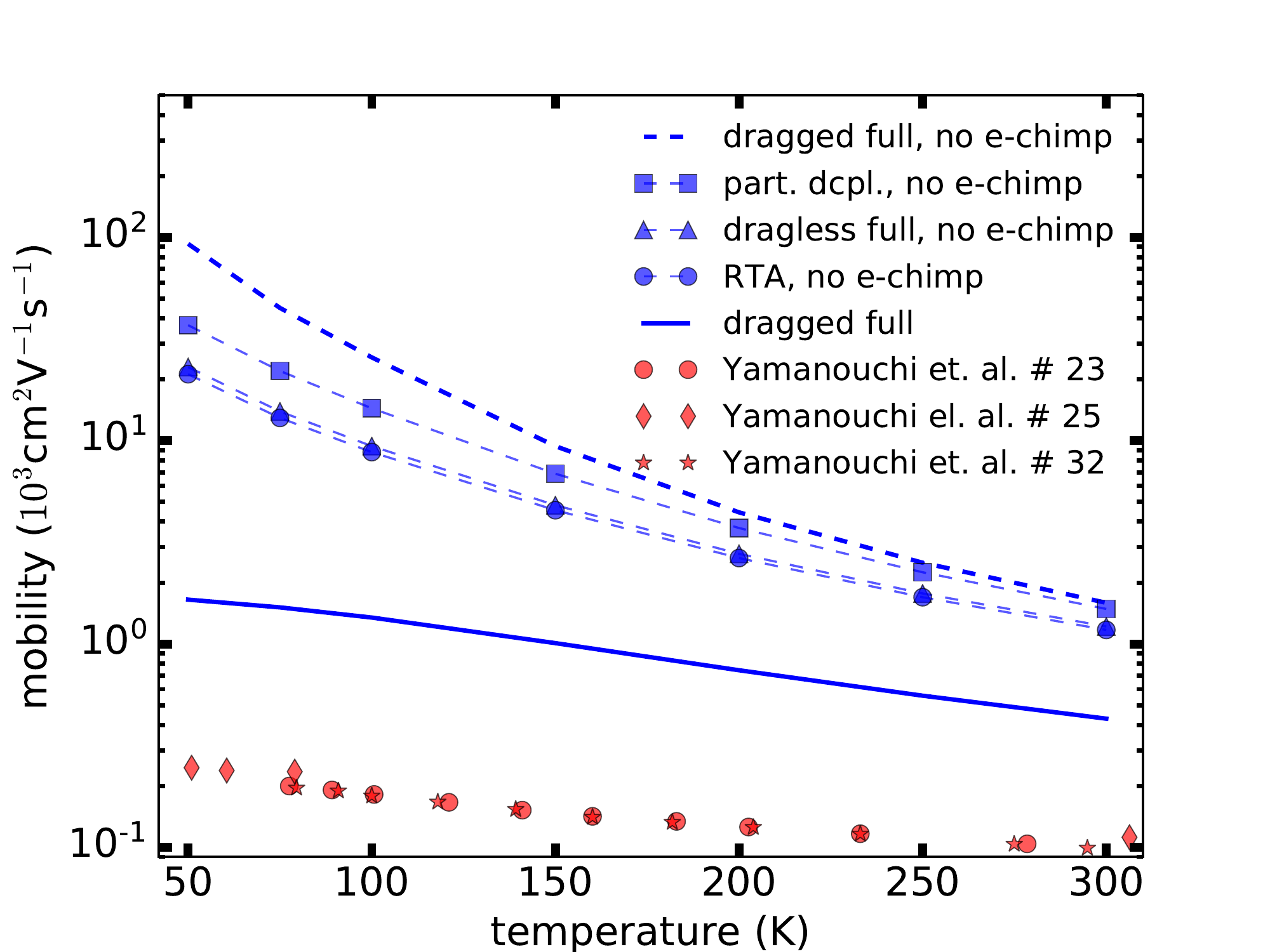}
	\caption{Temperature dependence of the mobility of silicon for an $n$-type carrier concentration of $2\times 10^{19}$ cm$^{-3}$. The red circles are measurements by Yamanouchi et. al. \cite{yamanouchi1967electric} on samples 23, 25, and 32, with dopant densities $2.28 \times 10^{19}$, $2.42 \times 10^{19}$, and $3.35 \times 10^{19}$ cm$^{-3}$, respectively.}
	\label{fig:highconc_mob_T}
\end{figure}

We consider first an $n$-type doped sample of low carrier density $2.75 \times 10^{14}$ cm$^{-3}$ over a range of temperatures. This is close to the carrier concentration in the best sample (number $537$) of Geballe and Hull's experimental work \cite{geballe1955seebeck}. We show in Fig. \ref{fig:lowconc_Q_T} a comparison of the calculated and the measured thermopower magnitudes. The experimental data is collected from Fig. 1 of Ref. \cite{geballe1955seebeck}. The black curve is the calculated total thermpower. In the Seebeck picture, this includes the phonon drag contribution. And in the Peltier picture, this is the sum of the electronic contribution and the phonon contribution, the latter being purely due to the electron drag effect. The Peltier picture also allows a clean separation of the electronic and the phonon contributions. This Peltier breakdown of the total thermopower is shown in Fig. \ref{fig:lowconc_Q_T}. The solid blue curve is the electronic contribution to the thermopower. This contribution decreases slightly with decreasing temperature. The solid green curve is the phonon thermopower. This starts off as significantly lower than the electronic thermopower at $300$ K, but quickly overtakes the latter around $175$ K, before dominating by more than an order of magnitude at $50$ K. The calculated total thermopower is in excellent agreement with the experimental measurements (red circles). Without the drag effect, the phonon contribution would be trivially zero, and the electronic contribution would be an order of magnitude off from the experimental values at low temperatures. Furthermore, without drag, the temperature dependence of the thermopower would be spectacularly wrong. The explanation of the origin of the strong drag effect in silicon has previously been given in Refs. \cite{zhou2015ab, fiorentini2016thermoelectric}, and is not reproduced here. These earlier \textit{ab initio} works captured the strong drag behavior using a partially decoupled solution of the e and ph BTEs. Nevertheless, they found excellent agreement with experimental measurements. Our calculations corroborate their finding that the partially decoupled solution is indeed sufficient to the capture the strong drag effect in silicon -- the green and blue squares denote the partially decoupled solutions, and they coincide nearly perfectly with corresponding full drag curves. For this case, the RTA and dragless full solutions of the e BTE also give essentially the same result and are not shown on the plot to reduce clutter. However, this is not guaranteed to hold true for all materials, and, in general, a full drag solution of the e-ph BTEs should be used.

Fig. \ref{fig:highconc_Q_T} shows the calculated thermopower of an $n$-type sample with a high density of $2.7 \times 10^{19}$ cm$^{-3}$, matching that of sample 140 in Ref. \cite{geballe1955seebeck}. Good agreement with measured data is again obtained. The percentage difference of the calculated total thermopower from the experimental value near $50$ ($300$) K is around  $15$ ($1$)\%. Note that the calculated thermopower without the phonon component again significantly underestimates the measured thermopower across the full range of temperatures considered.

\begin{figure}[h]
	\centering
	\includegraphics[width=0.8\linewidth]{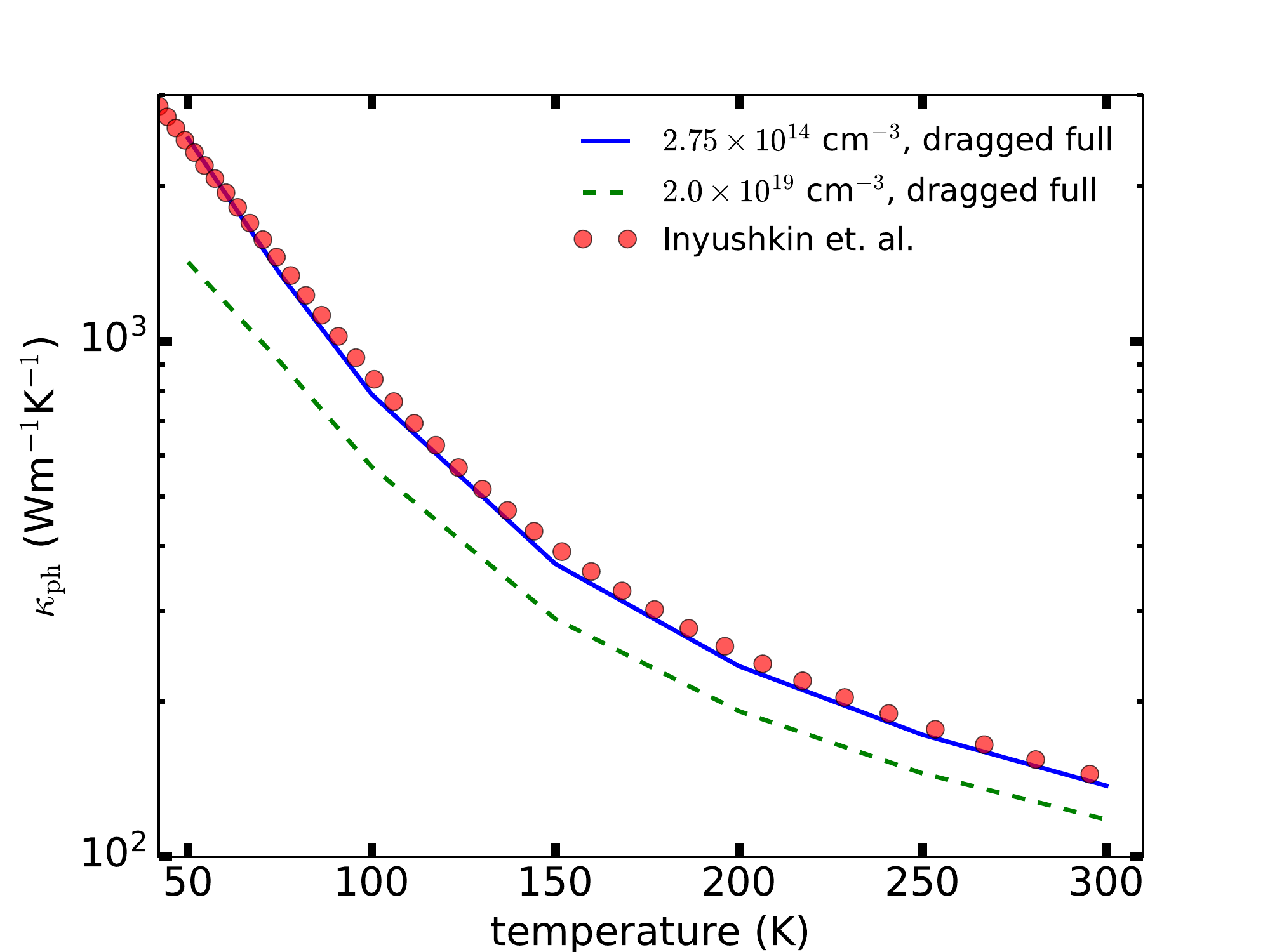}
	\caption{Temperature dependence of the phonon thermal conductivity of silicon for $n$-type carrier concentrations of $2.75\times 10^{14}$ and $2\times 10^{19}$ cm$^{-3}$. The red circles are measurements on high purity samples with natural isotopic mix \cite{inyushkin2004isotope}.}
	\label{fig:kph_T}
\end{figure}

In Fig. \ref{fig:lowconc_mob_T}, we present the calculated temperature dependent mobilities for a carrier concentration of $2.75\times 10^{14}$ cm$^{-3}$. Excellent agreement is found over the entire temperature range considered here with experimental measurements (shown in red circles) on similar low-doped samples \cite{canali1975electron}. The effect of phonon drag on this quantity is negligible. In fact, the difference between the dragged full, RTA, dragless full, and partially decoupled solutions is very small, and, to reduce clutter, we do not show the latter two results on the plot. This corroborates the finding in Ref. \cite{zhou2015ab}.

Fig. \ref{fig:highconc_mob_T} shows the calculated temperature dependent mobilities for a degenerate carrier concentration of $2\times 10^{19}$ cm$^{-3}$ and the comparison to experimental data from samples with similar electron concentrations from Ref. \cite{yamanouchi1967electric}. The calculated mobilities including e-chimp scattering (solid blue curve) are between a factor of 4 to 5 higher than the measured values (red circles). The RTA, dragless full, and partially decoupled solutions give nearly the same values as the dragged full solution and, to reduce clutter, we do not show these points on this plot. The large discrepancy between the calculated and the measured mobilities could be due to a combination of multiple reasons. First, the measurements were done on compensated samples and the acceptor and donor compositions were not reported in Ref. \cite{yamanouchi1967electric}. Our calculations have assumed that the mobile electron concentration is equal to the ionized donor density, while the acceptor density has been taken to be zero. In the actual samples, acceptors provide additional charge scattering centers thus lowering the mobility. Second, the e-chimp scattering employed in the calculation assumes a static, Thomas-Fermi screened Coulomb interaction treated in the Born approximation. It is known that the Thomas-Fermi model leads to a significant over-screening of the interaction in the degenerate limit \cite{chattopadhyay1981electron}. The validity of the Born approximation is also stretched in this limit and more sophisticated non-perturbative approaches might be better suited. Lastly, the consideration of the e-plasmon scattering, which is currently not included in our calculation, has been shown to reduce the mobility of silicon significantly at high carrier concentrations \cite{caruso2016theory}. A more rigorous treatment of the e-chimp scattering along with e-e scattering is planned for a future version of \texttt{elphbolt}.

We also show in Fig. \ref{fig:highconc_mob_T} the calculated mobility with the e-chimp interaction turned off (dashed blue curve). Here we envision that the mobile charge carriers are created in a region in which charged dopants do not exist, which could happen, for example, through modulation doping \cite{stormer1979two}, or in GaN/AlGaN heterostructures \cite{ambacher1999two}. In this case, the phonon drag effect leads to a large increase in the mobility -- at 300 (50) K, the drag gain of mobility is about a factor of 2.5 (50). Such high gains in the mobility can, potentially, be exploited along with those previously identified in the thermopower \cite{zhou2015ab, fiorentini2016thermoelectric, yalamarthy2019significant} to boost the thermoelectric figure-of-merit. Note that in this case the phonon drag effect is not fully captured by the partially decoupled solution which predicts about half the value given by the dragged full solution at 50 K. At this temperature, the RTA and dragless full solutions undershoot the value of the dragged full solution by about a factor of 5.

Finally, in Fig. \ref{fig:kph_T} we compare the calculated phonon thermal conductivities against measurements on high purity samples of natural silicon. The measured values (red circles) are taken from Ref. \cite{inyushkin2004isotope}. Remarkable agreement is found over the full temperature range considered for the low concentration case (solid blue curve). We also plot the results for a high concentration case in dashed green. The ph-e interactions cause a weaker suppression at high temperatures compared to that in the low temperature limit even at a high carrier concentration of $2\times 10^{19}$ cm$^{-3}$. This happens because the ph-ph scattering rates dominate over the ph-e ones at high temperatures. This is consistent with the findings in Ref. \cite{liao2015significant}. At low temperatues, the low energy acoustic phonons progressively contribute more to the thermal conductivity, while, at the same time, the ph-e scattering rates begin to dominate over the ph-ph ones for these modes. This results in a stronger suppression of the thermal conductivity at low temperatures for the high doped case. The electron drag effect on the phonon thermal conductivity is found to be small in this material, as has been shown earlier in Refs. \cite{zhou2015ab,fiorentini2016thermoelectric}. This has also been shown to be true for gallium arsenide \cite{protik2020coupled} and silicon carbide \cite{protik2020electron}. The reason for this has been discussed in these references. The dragged full, dragless full, partially decoupled, and the RTA solutions all give nearly the same results and only the first type is shown on the plot.

\section*{Discussion}\label{discussion}
In this work, we discussed the theory and implementation of \texttt{elphbolt} -- an efficient code for solving the coupled electron-phonon Boltzmann transport equations. This gives \textit{ab initio} access to the thermal, charge, and thermoelectric transport properties in materials, and the effect of e-ph drag on them. The code is distributed as Free/Libre software under the GNU General Public License version 3 that gives the user the right to use, modify, and distribute the original and their modified versions of the software. The code combines the object-oriented and the procedural programming styles, has a clean, modular structure, features \texttt{coarray} parallelization, and is well-documented. This makes the code easy to extend. For future releases, we plan to include a more rigorous treatment of impurity scattering, electron-electron interactions, provisions for including four-phonon interactions \cite{han2022fourphonon}, quadrupolar corrections, and magnetotransport.

\section*{Methods}\label{methods}
\subsection*{Calculation details}
We use the norm-conserving, Perdew-Zunger, local density approximation (LDA) pseudopotential \cite{perdew1981self}. A relaxed lattice constant of $5.40 \text{ \AA}$ is found. The phonon calculation is peformed using a $12\times 12\times 12$ $\kk$-mesh and a $6\times 6\times 6$ $\qq$-mesh. The \texttt{EPW} calculation is done with $4$ valence and $4$ conduction bands, using an initial guess of $sp^{3}$ projections on the two basis atoms. The IFC3s are calculated using a $5\times 5\times 5$ supercell ($250$ atoms) with a $6$ nearest neighbor cut-off and $\Gamma$-point sampling. In the transport calculation, we include e-ph, e-chimp (assuming singly charged impurities), ph-ph, ph-e, and ph-iso scattering. For the transport calculations, converged $50\times 50\times 50$ $\qq$- and $150\times 150\times 150$ $\kk$-meshes, denoted $(50, 150)$, are used. We use a \texttt{gcc} build of \texttt{elphbolt} with the \texttt{OpenCoarrays} library \cite{fanfarillo2014opencoarrays} for \texttt{coarray} support.

A typical calculation of the fully coupled e-ph BTEs using the $(50,150)$ mesh set takes on the order of $3000$ cpu-hours. A breakdown of the various important components is given in Table \ref{tab:computetime}. These numbers were calculated on 4 nodes each equipped with 28 Intel(R) Xeon(R) CPU E5-2680 v4 @ 2.40GHz cores for an $n$-type carrier concentration of $2.75\times 10^{14}$ cm$^{-3}$ at $300$ K and a relative convergence threshold of $0.0001$. The coupled e-ph BTEs required 6 iterations to converge, whereas the decoupled e and ph BTEs took 4 and 7 iterations, respectively. The total run time, of course, can vary significantly depending on the speed of disk read/write and that of the cpus, the temperature, and the carrier concentration. Note that once the interaction vertices are calculated, these can then be reused during the solutions of the BTEs for different temperatures and carrier concentrations.

\begin{table}
	\centering
	{\begin{tabular}{lccccc}
	\hline
	task & e-ph vertex & ph-ph vertex & e-ph BTEs & e BTE & ph BTE \\
	time (cpu-hours) & 744	& 545 & 1467 & 66 & 28\\
	\hline
	\end{tabular}}
	\caption{Breakdown of the computational time needed for the various expensive parts of the code for silicon. An $n$-type carrier concentration of $2.75\times 10^{14}$ cm$^{-3}$ at $300$K using a $(50, 150)$ mesh set is considered here.}
	\label{tab:computetime}
\end{table}

\section*{Data availability statement}
The input files needed to generate both the force constants and Wannier space data required to reproduce the results in this work are available from github \cite{protik_elphbolt_2021}.

\section*{Code availability statement}
The code used in this work is available from github \cite{protik_elphbolt_2021}.

\section*{Acknowledgments}
This project was funded by the EU-H2020 through H2020-NMBP-TO-IND project GA n.814487 (INTERSECT). ICN2 is supported by the Severo Ochoa program from Spanish MINECO (Grant No. SEV-2017-0706) and the CERCA Program of Generalitat de Catalunya. Work at Boston College (contributions to code testing and \textit{ab initio} thermoelectric transport calculations for silicon) was supported by the U.S. Department of Energy (DOE), Office of Science, Basic Energy Sciences under award \# DE-SC0021071. NHP acknowledges helpful discussions with Vladimir Dikan, Jos\'{e} Mar\'{i}a Escart\'{i}n, Xavier Cartoix\`{a}, and Riccardo Rurali. We thankfully acknowledge the computer resources at MareNostrum and LaPalma and the technical support provided by Barcelona Supercomputing Center (FI-2021-1-0016) and Center for Astrophysics in La Palma (QS-2021-1-0022), respectively. We also acknowledge computational support from the Boston College Linux clusters and those at ICN2 provided by Grant PGC2018-096955-B-C43 funded by MCIN/AEI/10.13039/501100011033 and “ERDF A way of making Europe”.

\section*{Author contributions}
NHP is the developer and maintainer of the \texttt{elphbolt} code. NHP performed some of the \textit{ab initio} calculations and wrote the first draft of the manuscript in consultation with and under the supervision of MP and PO. CL performed some of the \textit{ab initio} calculations under the supervision of DB. All authors took part in the preparation of the manuscript.

\section*{Competing interests}
The authors declare no competing interests.

\bibliographystyle{naturemag}
\bibliography{refs_abbrv}


\nolinenumbers

\end{document}